\documentclass[a4paper,11pt]{article}
\usepackage{pos}

\usepackage{etex, xy}
\xyoption{all}

\usepackage{graphicx}

\newcommand{\xdownarrow}[1]{%
	{\left\downarrow\vbox to #1{}\right.\kern-\nulldelimiterspace}
}

\newcommand{\fl}[1]{\text{\scriptsize$\big\lfloor\tfrac{#1}2\big\rfloor$}}

\renewcommand{\hookAfterAbstract}{%
	\par\bigskip
	\textsc{DESY-24-121}\\
	\textsc{RISC Report number 24-05}\\
	PoS (LL2024) 031
}

\newcommand{\KK}{\mathbb{K}}

\newcommand{\ZZ}{\mathbb{Z}}

\newcommand{\ep}{\varepsilon}

\newcommand{\Cc}{\mathcal C}

\newcommand{\K}{{\mathcal K}}

\title{
Challenges for analytic calculations of the massive three-loop form factors}
\ShortTitle{Challenges for analytic calculations of the massive three-loop form factors}

\author[a]{J. Bl\"umlein}
\author*[b]{A. De Freitas}
\author[a]{P. Marquard}
\author[b]{C. Schneider}

\affiliation[a]{Deutsches Elektronen-Synchrotron DESY, Platanenallee 6, 15738 Zeuthen, Germany}
\affiliation[b]{Johannes Kepler University Linz, Research Institute for Symbolic Computation (RISC), \\ 
Altenbergerstraße 69, A–4040, Linz, Austria}

\emailAdd{johannes.bluemlein@desy.de}
\emailAdd{afreitas@risc.jku.at}
\emailAdd{peter.marquard@desy.de}
\emailAdd{carsten.schneider@risc.jku.at}

\abstract{The calculation of massive three-loop QCD form factors using in particular the large moments method has been successfully applied to quarkonic contributions in \cite{Blumlein:2023uuq}. We give a brief review of the different steps of the calculation  and report on improvements of our methods that enabled us to push forward the calculations of the gluonic contributions to the form factors.}

\FullConference{Loops and Legs in Quantum Field Theory (LL2024)\\
 14-19, April, 2024\\
Wittenberg, Germany\\}

\begin{document}
\maketitle

\section{Introduction}

Massive QCD form factors constitute basic building blocks for several important processes in particle physics, such as heavy quark production at $e^+e^-$ colliders and Higgs or gauge boson decay into heavy quarks.
In order to match the precision of available experimental results, these form factors must be computed at higher orders in the corresponding perturbative expansions.
The two-loop contributions were computed up to $O(\ep^0)$ about two decades ago in Refs. \cite{Bernreuther:2004ih, Bernreuther:2004th, Bernreuther:2005rw, Bernreuther:2005gw}, 
where $\ep$ is the dimensional regularization parameter, which is given by $\ep=(4-D)/2$, where $D$ is the dimension. 
This calculation was later extended to $O(\ep)$ in \cite{Gluza:2009yy} and to $O(\ep^2)$ in \cite{Ablinger:2017hst}. 
These higher orders in $\ep$ are needed in order to combine them with the corresponding higher loop order calculations.
The first analytic results at three loops were obtained in the planar limit in Refs. \cite{Henn:2016kjz,Henn:2016tyf,Ablinger:2018yae}.
The high energy behaviour of three loop form factors was studied in Refs. \cite{Ahmed:2017gyt,Blumlein:2018tmz}, 
and light-fermion contributions were computed in Refs. \cite{Lee:2018nxa,Lee:2018rgs}.
Analytic results in terms of harmonic polylogarithms \cite{Remiddi:1999ew} for the solvable parts of quarkonic contributions 
(that is, contributions from diagrams containing a closed fermion loop) were given in \cite{Blumlein:2019oas}.
More recently, numerical results for all form factors, including singlet and anomaly contributions, were obtained in Refs. \cite{Fael:2022miw,Fael:2022rgm,Fael:2023zqr}.
Later, in Ref. \cite{Blumlein:2023uuq}, we used the large moments method \cite{Blumlein:2017dxp,Ablinger:2018zwz,SolveCoupledSystem} 
to obtain analytic results for the quarkonic contributions to all non-singlet form factors.
In these proceedings, we report on the progress using this method to compute the gluonic contributions, that is, the contributions stemming from diagrams with no internal closed fermion loops.

The large moments method derives its name from the fact that it was first applied to the computation of the factorizable parts of massive three-loop operator matrix elements and related problems using a large number of Mellin moments~\cite{Ablinger:2022wbb,Ablinger:2023ahe,Ablinger:2024xtt,Blumlein:2022gpp,Blumlein:2021enk,Blumlein:2009tj}. 
This method can also be applied if we have a large number of expansion coefficients, instead of Mellin moments, for a physical quantity depending on a single variable. 
We will give a brief review of this method in Section \ref{sec:review} and focus on concrete challenges of the large moments method related to the form factor project in Section~\ref{Sec:ChallengesLM}.
In Section \ref{sec:gluonic} we will present the progress we have made applying this method to the gluonic case, which is considerably more
involved than the quarkonic one. In Section \ref{sec:conclusions}, we give the conclusions.

\section{Review of the method}
\label{sec:review}

The three-point functions with an external massive quark-anti-quark pair coupling to a vector, axial-vector, scalar or pseudo-scalar boson are given (in that order) by
\begin{eqnarray}
\Gamma_V^{\mu} &=& \gamma^{\mu} F_{V,1}\left(\hat{s}\right) - \frac{i}{2 m} \sigma^{\mu\nu} q_{\nu} F_{V,2}\left(\hat{s}\right)\, , \label{eq:FFvec} \\
\Gamma_A^{\mu} &=& \gamma^{\mu} \gamma_5 F_{A,1}\left(\hat{s}\right) - \frac{1}{2 m} q^{\mu} \gamma_5 F_{A,2}\left(\hat{s}\right)\, , \\
\Gamma_S &=& m F_{S}\left(\hat{s}\right)\, , \\
\Gamma_P &=& i m F_{P}\left(\hat{s}\right)\,; \label{eq:FFpse}
\end{eqnarray}
here $\hat{s}=q^2/m^2$, where $q$ is the momentum of the boson and $m$ is the heavy quark mass. The functions $F_{V,1}$,
$F_{V,2}$, $F_{A,1}$, $F_{A,2}$, $F_S$ and $F_P$ are the so called form factors, which can be extracted from Eqs. (\ref{eq:FFvec}-\ref{eq:FFpse}) by applying suitable projectors.
The first steps of our calculation are standard: We generated the Feynman diagrams using {\tt QGRAF} \cite{Nogueira:1991ex}, 
and after introducing the aforementioned projectors, 
we performed the Lorentz algebra using {\tt FORM} \cite{Vermaseren:2000nd,Tentyukov:2007mu} and the color algebra using {\tt Color} \cite{vanRitbergen:1998pn}. 
After this, we end up with expressions for the form factors in terms of linear combinations of scalar Feynman integrals, 
which we reduced to master integrals using {\tt Crusher} \cite{CRUSHER}. At this point, a system of first order differential equations obeyed by the master integrals can be obtained.
In many similar perturbative calculations,
the corresponding systems of differential equations can be solved in terms of closed form solutions involving harmonic polylogarithms and their generalizations \cite{Vermaseren:1998uu,Blumlein:1998if,Ablinger:2011te,Ablinger:2013cf,Ablinger:2014bra,Remiddi:1999ew,Blumlein:2003gb,Ablinger:2013hcp,Ablinger:2014rba}.
Unfortunately, this is not always possible, which is often the case in calculations involving massive particles, such as the present one. 
If we try to decouple the system using, e.g., the package \texttt{OreSys}~\cite{OreSys}, we find out that the resulting higher order differential equations are
not first order factorizable \cite{nonfact}, 
leading to complicated solutions involving elliptic or even higher trascendental functions, as well as iterated integrals over these functions, 
which can be difficult to use for numerical evaluations and phenomenological purposes. For this reason, one is forced to adopt a different tactic. 
In Refs. \cite{Fael:2022miw,Fael:2022rgm,Fael:2023zqr}, the authors tackled the problem by using the differential equations satisfied by the master integrals
in order to find approximate numerical solutions to the master integrals consisting
of power and power-log expansions with numerical coefficients obtained by matching the solutions at different values of $\hat{s}$. 
In Ref.~\cite{Blumlein:2023uuq}, we used a strategy also based on differential equations, but instead of using the differential equations obeyed by the master integrals.
Using guessing algorithms \cite{GSAGE}, 
we derived recursion relations and differential equations from the sequences of rational numbers multiplying the constants appearing in the expansion at $x=1$ of the form factors,
where,
\begin{equation}
\hat{s} = -\frac{(1-x)^2}{x}\, . \label{eq:s2x}
\end{equation}

The form factors can be written as follows
\begin{equation}
F_I(x) = \sum_{k,l} \Cc_k \K_l F_{I;k,l}(x),
\label{eq:FF-color-const}
\end{equation}
where $I$ labels any of the possible form factors, the $\K_l$'s are any of the constants in the following list
\begin{equation}
\K_l \in \left\{1, \, \zeta_2, \, \zeta_3, \, \ln(2) \zeta_2, \, \zeta_2^2, \, \ln^2(2) \zeta_2, \, {\rm Li}_4\left(\textstyle{\frac{1}{2}}\right), \, \ln^4(2), \, \zeta_2 \zeta_3, \, \zeta_5\right\},
\label{eq:constants1}
\end{equation}
where the $\zeta_k = \zeta(k)$, i.e., the Riemann $\zeta$ function evaluated at integer values, and the $\Cc_k$'s are color factors, which at three loops can take any of the following values
\begin{eqnarray}
	\Cc_k &\in&  \left\{C_A^2 C_F, \, C_A C_F^2, \, C_F^3, \, n_l T_F C_F^2, \, n_l T_F C_F C_A, \, n_l^2 T_F^2 C_F, \, n_l n_h T_F^2 C_F, \, n_h^2 T_F^2 C_F, \right. \nonumber \\ && \left.
	\phantom{\left\{ \right.}
	n_h T_F C_F^2, \, n_h T_F C_F C_A
	\right\},
	\label{eq:color-factors-and-n}
\end{eqnarray}
with
\begin{equation}
C_A = N_c, \quad C_F = \frac{N_c^2-1}{2 N_c}, \quad T_F = \frac{1}{2},
\end{equation}
where $N_c$ is the number of colors, while $n_l$ and $n_h$ are the number of light and heavy quarks, respectively. 
The first three color factors in (\ref{eq:color-factors-and-n}) are the ones appearing in the gluonic case, while the remaining ones correspond to the quarkonic case.

The split in terms of the constants in Eq. (\ref{eq:FF-color-const}), is done in such a way
that the functions $F_{I;k,l}(x)$ can be expanded around $x=1$ (which corresponds to $\hat{s}=0$) as follows
\begin{equation}
F_{I;k,l}(x) = \sum_{i=0}^{\infty} r_{I;i}^{(k,l)}(0) y^i,
\label{eq:s=q-expansion}
\end{equation}
where $y=1-x$, and the $r_{I;i}^{(k,l)}(0)$ are rational numbers. Alternatively, we may also work with expansions around $\hat{s}=0$, in which case, we will have equations
(\ref{eq:FF-color-const}) and (\ref{eq:s=q-expansion}) with $x$ and $y$ replaced by $\hat{s}$, 
and with different sequences of rational numbers in (\ref{eq:s=q-expansion}).
In the quarkonic case, and more specifically, in the case of the last two color factors in (\ref{eq:color-factors-and-n}),
we were able to generate up to 8000 expansion coefficients by using the large moments method; for details we refer to Section~\ref{Sec:ChallengesLM}.
Using this information, we follow the guess and solve strategy that has been exemplified for the first time rigorously in particle physics in~\cite{Blumlein:2009tj}. 
More precisely, using the 8000 coefficients, we succeeded in guessing recursion relations as well as differential equations associated with these sequences of rational numbers using the {\tt Sage} package {\tt ore\_algebra} \cite{GSAGE}.
These recursions and differential equations were solvable for all but three of the constants in (\ref{eq:constants1}), namely,
\begin{equation}
\{1, \, \zeta_2, \, \zeta_3\} \label{eq:constants2}.
\end{equation}

In all other cases, we solved the recursions by using the summation package \texttt{Sigma}~\cite{Sigma} containing the algorithms introduced in~\cite{ABPS:21,Schneider:23} and the references therein. Finally, we performed the infinite sum for the corresponding constants using \texttt{Sigma} and the package~\texttt{HarmonicSums}~\cite{Vermaseren:1998uu,Blumlein:1998if,Ablinger:2011te,Ablinger:2013cf,Ablinger:2014bra,Remiddi:1999ew,Blumlein:2003gb,Ablinger:2013hcp,Ablinger:2014rba}.
In the case of the constants in (\ref{eq:constants2}), since we cannot solve the corresponding recursions, we proceeded as follows:
\begin{enumerate}
\item We used the recursion relations to generate a much larger number of expansion coefficients for the expansion at $x=1$. 
This allows us to evaluate this expansion with very high precision even at values very close to $x=0$.
\item An ansatz in terms of a power-log expansion at $x=0$ was inserted in the differential equation for each constant in (\ref{eq:constants2}). 
This allowed us to express all coefficients of the expansion at $x=0$ in terms of only a few of them (as many as the order of the differential equation).
\item The expansions at $x=1$ and $x=0$ were matched at a point close to zero, and the coefficients left from the previous step were evaluated numerically with very high precision,
making it possible to determine all of them in terms of known constants using the PSLQ algorithm \cite{PSLQ,Bailey:1999nv}.
\item The same procedure was repeated for the expansions at the 2-particle and the 4-particle thresholds, albeit in these cases we worked with the expansions at $\hat{s}=0$, 
and the expansions at $\hat{s}=4$ and $\hat{s}=16$ were given in terms of the variables $\sqrt{4-\hat{s}}$ and $\sqrt{16-\hat{s}}$, respectively. 
In the case of the 2-particle threshold, we also used the PSLQ algorithm to determine the coefficients, but new constants needed to be introduced, which were chosen among a few of
the coefficients themselves.
\end{enumerate}
We followed these steps for all of the form factors, allowing us to cover the full kinematic range in $\hat{s}$. In Ref. \cite{Blumlein:2023uuq}, we even went a bit further, 
and after combining all the results for all the constants in (\ref{eq:FF-color-const}), we repeated what we did with the sequences of rational numbers associated to
the expansions at $x=1$, but now with the resulting sequences of rational numbers in the power-log expansions at $x=0$, 
and again obtained the corresponding recursions relations, several of which could be solved and summed with {\tt Sigma}, leading to new analytic results at that expansion point.


\section{Challenges arising in the large moments method}\label{Sec:ChallengesLM}
In the following we present the basic idea of the large moments method~\cite{Blumlein:2017dxp} and elaborate on improvements that supplement the ideas given in~\cite{Ablinger:2018zwz,Blumlein:2019oas,SolveCoupledSystem,Blumlein:2023uuq}.
Starting with a physical expression\footnote{In the moment we suppress the dependence on the dimensional parameter $\ep$.} $\hat{P}(x)$ in terms of Feynman integrals, the IBP approach~utilizes integration by parts (IBP) methods~\cite{Chetyrkin:1981qh,Laporta:2001dd,CRUSHER} and returns an alternative representation
\begin{equation}\label{Equ:HatPExpr}
	\hat{P}(x)=q_1(x)\hat{I}_1(x)+\dots+q_i(x)\hat{I}_i(x)+\dots+q_{\lambda}(x)\hat{I}_{\lambda}(x)
\end{equation}
in terms of so-called master integrals $\hat{I}_i(x)$ and rational functions $q_i(x)\in\KK(x)$ where $\KK=\KK'(\ep)$ is a rational function field over a field $\KK'$ containing the rational numbers as subfield; here $x$ stands for any variable and can take over the role of $x$, $y:=1-x$ or $\hat{s}$ as introduced in~\eqref{eq:s2x}. While the input expression may contain millions of such integrals, the output consists of much less integrals; for a concrete calculation of the form factor project one gets, e.g., $\lambda=2506$ such master integrals. In addition, one obtains a coupled system of first-order linear differential equations in terms of the unknown master integrals. Under the assumption that the master integrals $\hat{I}_i(x)$ have a power series representation (or Laurent series representation) in $x$, 
we can apply our large moments method, that can be split into two parts:
\begin{enumerate}
	\item Compute a large number $\nu$ of values, say $\nu=8000$ for the quarkonic or $\nu=20000$ for the gluonic case, of each master integral, i.e., compute 
	$I_{i}(n)$ for $n=0,\dots,\nu$ s.t.
	\begin{equation}\label{Equ:hatIExp}
		\hat{I_i}(x)=\sum_{n=0}^{\nu}I_{i}(n)x^n+O(x^{\nu+1}).
	\end{equation}
	\item Plug these expansions~\eqref{Equ:hatIExp} into~\eqref{Equ:HatPExpr} and derive the first $\nu$ coefficients of the series expansion 
	\begin{equation}\label{Equ:PExpansion}
		\hat{P}(x)=\sum_{n=0}^{\nu}P(n)x^n+O(x^{\nu+1}).
	\end{equation}
	
\end{enumerate}
\textit{Remark:} Since $\hat{I_i}(x)$ and $\hat{P}(x)$ depend also on the dimensional parameter $\ep$, also $I_{i}(n)$ and $P(n)$ depend on it. In a naive approach one may look for $P(n)$ and $I_i(n)$ in $\KK=\KK'(\ep)$. However, in applications from particle physics, one seeks not only for an expansion in $x$ but also in $\ep$. For the three-loop case, the $\ep$-expansion of $\hat{P}(x)$ usually starts at $1/\ep^3$ and one is interested in the coefficients up to the constant term.

\medskip

In step 1 uncoupling algorithms implemented in the package \texttt{OreSys}~\cite{OreSys} are used to obtain scalar linear differential equations in terms of the unknown master integrals $\hat{I}_i$ of the form
\begin{equation}\label{Equ:DHatI}
	\alpha_0(x) \hat{I}_i(x)+\alpha_1(x) D_x \hat{I}_i(x)+\dots+\alpha_{\rho}(x) D_x^{\rho}\hat{I}_i(x)=\hat{r}(x)
\end{equation}
where $\rho\in\ZZ_{\geq0}$ is relatively small, say$\leq 15$, the $\alpha_i(x)\in\KK[x]$ are polynomials in $x$ and $\hat{r}(x)$ is given in terms of other master integrals like~\eqref{Equ:HatPExpr}; more precisely, $\hat{r}(x)$  is again of the form~\eqref{Equ:HatPExpr} where one can derive the first $\nu$ coefficients of the $x$-expansions of the occurring master integrals either by different methods, like symbolic summation~\cite{Sigma,Sagex} or by the large moments method applied recursively. Given these expansions, we compute (as in step 2 described in more details below)  the $x$-expansion of $\hat{r}(x)$, i.e.,
$$\hat{r}(x)=\sum_{n=0}^{\nu}r(n)x^n+O(x^{\nu+1}).$$
Then by plugging in~\eqref{Equ:hatIExp} into~\eqref{Equ:DHatI} and performing coefficient comparison w.r.t.\ $x^n$ one obtains a linear recurrence of the form
\begin{equation}\label{Equ:Rec}
	\beta_0(n)I_{i}(n)+\beta_1(n)I_{i}(n+1)+\dots+\beta_{u}(n)I_{i}(n+u)=r(n)
\end{equation}
with polynomials $\beta_i(x)\in\KK[x]$ and $u\in\ZZ_{\geq0}$. 
Given this representation and the first $u$ initial values (coming from other methods like symbolic summation or integration~\cite{Sagex}), one can now compute efficiently (in linear time) the coefficients $I_{i}(0),\dots,I_{i}(\nu)$. 

\medskip

\noindent\textit{Remark:} Note that $\KK=\KK'(\ep)$, i.e., the coefficients $\beta_i$ also depend on $\ep$. Using the algorithms described in~\cite{BKSF12,Blumlein:2017dxp} one can derive in addition the desired $\ep$-expansions of the coefficients $I_i(n)$ up to the desired order. 
In~\cite{Blumlein:2017dxp,Ablinger:2018zwz,Blumlein:2019oas,SolveCoupledSystem,Blumlein:2023uuq}  we address various aspects of step 1 to reduce the orders $\rho$ and $u$ in~\eqref{Equ:DHatI} and~\eqref{Equ:Rec}, respectively, and elaborate on various tactics to extract the $\ep$-expansions in combination with the different uncoupling algorithms from~\cite{OreSys}. 

\medskip

In the following we will focus on step 2 of the method and present various improvements to gain significant speed-ups that are relevant to tackle the gluonic and quarkonic form factors. As indicated aboe, these improvements play also a crucial role to derive the $x$-expansion (and $\ep$-expansion) of $\hat{r}(x)$ in step~1 efficiently.

Recently, we dealt in our gluonic form factor calculations with huge expressions of the form~\eqref{Equ:HatPExpr}. For a typical example we got an expression with $\lambda=2506$ master integrals which required about 106 GB memory in Mathematica. As it turns out, the expressions $q_i(x)$ itself are split into many smaller rational functions, say $\mu=1213$ elements. One option would be to compute an $x$-expansion for each subexpression which would result in total to around a million expansions. To reduce the number of such expansions significantly, we first merged each $q_i(x)$ to one rational function by the divide-and-conquer strategy illustrated in Fig.~\ref{fig:MyTogether}.
\begin{figure}
$$q_i(x)$$
$$||$$
$$\underbrace{\frac{a_1(x)}{b_1(x)}+\dots+\frac{a_{\fl{\mu}}(x)}{b_{\fl{\mu}}(x)}}_{}+\underbrace{\frac{a_{\fl{\mu}+1}(x)}{b_{\fl{\mu}+1}(x)}+\dots+\frac{a_{\mu}(x)}{b_{\mu}(x)}}_{}$$

\vspace*{-0.9cm}

$$\hspace*{0.2cm}\xdownarrow{0.25cm}\hspace*{3.9cm}\xdownarrow{0.25cm}\hspace*{0.4cm}$$

\vspace*{-0.6cm}

$$\hspace*{2.2cm}\vdots\hspace*{1cm}\texttt{MyTogether}\hspace*{1.1cm}\vdots\hspace*{1cm}\txt{Divide \&\\ Conquer}$$

\vspace*{-0.6cm}

$$\hspace*{0.2cm}\xdownarrow{0.25cm}\hspace*{3.9cm}\xdownarrow{0.25cm}\hspace*{0.4cm}$$
$$\frac{A_1(x)}{B_1(x)}\hspace*{1.3cm}+\hspace*{1.3cm}\frac{A_2(x)}{B_2(x)}$$
$$||$$
$$\frac{A_1(x)B_2(x)+A_2(x)B_1(x)}{B_1(x)B_2(x)}$$
$$\hspace*{1.7cm}||\texttt{ MyCancel}$$
$$\hspace*{3cm}\frac{A(x)}{B(x)}\quad \quad\gcd(A,B)=1$$
\caption{\texttt{MyTogether}}
\label{fig:MyTogether}
\end{figure}
While the combining step in the inner recursions work rather efficiently using Mathematica (i.e., \texttt{MyCancel} is replaced by Mathematica's command \texttt{Cancel} or \texttt{Together}), in the outermost calls the expressions get very big. An extra complication is that the arising objects are not only rational functions in $x$ but also depend on the dimensional parameter $\ep$. This makes the gcd calculations extremely costly. 
In the final step of such a calculation one has to deal, e.g., with the situation
$$\frac{\overbrace{A_1(x,\ep)B_2(x)+A_2(x,\ep)B_1(x,\ep)}^{A'(x,\ep)=}}{\underbrace{B_1(x,\ep)B_2(x,\ep)}_{B'(x,\ep)}}\hspace*{1cm}\txt{$\deg_x(A')\leq 1422$\\
	$\deg_x(B')\leq 1405$.}$$
Then the challenge is to remove common factors in $A'$ and $B'$ in reasonable time. In our setting, it turns out that $B'(x,\ep)$ factors sufficiently nice. E.g, we obtain factorizations of the form
\begin{equation}\label{Equ:DenFactor}
	B'(x,\ep)=\text{factor}_1(x,\ep)\text{factor}_2(x,\ep)\dots\text{factor}_{71}(x,\ep),
\end{equation}
where the first factors are linear and the most complicated irreducible factors are of the form 
\begin{align*}
	&48 x^4\ep^6-9792 x^3 \ep^6+19296 x^2 \ep^6-19008 x \ep^6+9504\ep^6+1006 x^4 \ep^5+9264 x^3 \ep^5\\
	&-16896 x^2 \ep^5+15264 x\ep^5-7632 \ep^5-686 x^4 \ep^4-4136 x^3 \ep^4+7352 x^2 \ep^4\\
	&-6432x \ep^4+3216 \ep^4+95 x^4 \ep^3+2416 x^3 \ep^3-6664 x^2\ep^3+8496 x \ep^3\\
	&-4248 \ep^3-32 x^4 \ep^2-1270 x^3 \ep^2+4438x^2 \ep^2-6336 x \ep^2+3168 \ep^2\\
	&+25 x^4 \ep+330 x^3 \ep-1266 x^2\ep+1872 x \ep-936 \ep-4 x^4-32 x^3+128 x^2-192 x+96.
\end{align*}
In order to detect candidates that may cancel, we evaluated $A'(x,\ep)$, e.g., at $\ep=1234$ and use the fact that
$$\text{factor}(x,{1234}) \nmid A(x,{1234})\quad\Longrightarrow\quad \text{factor}(x,\ep) \nmid A(x,\ep).$$
Thus we may exclude all factors that cancel by performing much more efficient operations in $\KK'[x]$ instead of $\KK'[x,\ep]$. For the remaining candidates we carry out polynomial division
$$A(x,\ep)=q(x,\ep)\,\text{factor}(x,\ep)+r(x,\ep).$$
If $r(x,\ep)=0$, we remove the factor from $B$ and replace $A$ by $q$. If $r(x,\ep)\neq 0$ (which never happened), we keep the factor. Performing our divide and conquer strategy \texttt{MyTogether} with this cancellation tactic \texttt{MyCancel} we could compress the original expression $\hat{P}(x)$ from $106$ GB to $5.7$ GB. Note: (1) Applying the Mathematica command \texttt{Together} or \texttt{Cancel} to such a reduced expression $q_i(x)$ where the numerator and denominator are already  coprime, it takes around 10 hours that Mathematica observes that nothing can be canceled. In a nutshell, utilizing the specific structure that the denominators factor nicely (see~\eqref{Equ:DenFactor}) was essential to carry out the compactification in reasonable time.
(2) As elaborated in~\cite{MCA} there are many advanced algorithms available to perform rational function arithmetic for general problems. But for our concrete situation the above approach seems to be rather optimal.

Finally, we take each simplified term in~\eqref{Equ:HatPExpr} and compute the expansions
\begin{equation}\label{Equ:FracExp}
	\frac{A(x)}{B(x)}\,\hat{I}(x)=\sum_{n=0}^{\nu}c(n)x^n+O(x^{\nu+1})
\end{equation}
in parallel to get the final expansion~\eqref{Equ:PExpansion}.  Due to step~1 we may assume that we are given the expansion~\eqref{Equ:hatIExp} in $x$ where the coefficients $I_i(n)$ are also expanded up to a certain order in $\ep$. Thus in order to obtain the expansion in~\eqref{Equ:FracExp}, we need to compute the expansion of $\frac{A(x)}{B(x)}$ in $x$ (and also in $\ep$). In our three-loop case the expansion of the master integrals start at order $1/\ep^3$ und thus the $\ep$-expansion of $\frac{A(x)}{B(x)}$  has to be computed only up order $\ep^3$ to derive the desired constant term $\ep^0$ in~\eqref{Equ:FracExp}.
This task can be done by simple Taylor expansion using the Mathematica command \texttt{Series}.
Afterwards we carry out the expansion in $x$ to the few coefficients of the $\ep$-expansion. In short, it remains to compute an $x$-expansion of expressions of the form~\eqref{Equ:FracExp} with $A(x),B(x)\in\KK'[x]$ being free of $\ep$ and where the first $\nu+1$ coefficients $I_i(n)\in\KK'$ in~\eqref{Equ:hatIExp} (free of $\ep$) are given explicitly. As it turns out, the standard command \texttt{Series} is hopeless for large values $\nu$. As a consequence we separate the task and look first for an expansion of 
\begin{equation}\label{Equ:1OverB}
	\frac1{B(x)}=\sum_{n=0}^{\nu}h(n)x^n+O(x^{\nu+1}).
\end{equation}
Here we may use the classical result (see, e.g., \cite{KP11}) that the sequence $h(n)$ can be computed by a recurrence with constant coefficients that is determined by the denominator $B(x)$. However, a typical example is
\begin{align*}
	B(x)=&\overbrace{
		(x-2)^{26} (x-1)^{19} x^{32} (x+1)^2 (2 x-1)^2}^{=F_{\text{lin}}(x)}\times\\
	\times&(x^2-22 x+22)^9 (x^2-15x+15)^{10} (x^2-6 x+6)^9 (x^2-5 x+5)^{10}(x^2-2x+2)^{21}\\[-0.1cm]
	&(x^2-x+1)^{19} (x^2+x-1)^{10} (x^2+2 x-2)^{10} (x^2+3 x-3)^{11} (x^2+4 x-4)^9\\[-0.1cm]
	& (2 x^2+3 x-3)^9 (3 x^2-14x+14)^{12} (3 x^2-10 x+10) (3 x^2-8x+8)^{11}\\[-0.1cm]
	&(3 x^2-4x+4)^8 (3 x^2-2 x+2)^9 (3 x^2+x-1)^8 (3 x^2+2 x-2)^{10}(4 x^2+3 x-3)^9\\[-0.1cm]
	& (5 x^2-18 x+18)^{10} (5 x^2-16 x+16)^{11}(5 x^2-2 x+2)^9 (5 x^2+12 x-12)^7\\[-0.1cm]
	& (7 x^2-6 x+6)^9 (11x^2+2 x-2)^9 (27 x^2+32 x-32)^9 (29 x^2-2 x+2)^8\\[-0.1cm]
	& (99 x^2-238x+238)^{10} (x^4-8 x^3+28 x^2-40 x+20) (x^4-6 x^3+18 x^2-24 x+12)^8\\[-0.1cm] 
	& (x^4-4 x^3+5 x^2-2 x+1)^9 (x^4+8 x^3-32 x^2+48 x-24)^8 (3 x^4+2x^3-6 x^2+8 x-4)^8\\[-0.1cm]
	& (5 x^4-152 x^3+272 x^2-240 x+120)^{10} (5 x^4-29x^3+27 x^2+4 x-2)^9\\[-0.1cm]
	& (5 x^4+16 x^3-40 x^2+48 x-24)^{12} (5 x^4+184 x^3-352x^2+336 x-168)^9\\[-0.1cm]
	& (7 x^4-31 x^3+25 x^2+12 x-6)^{10} (9 x^4-43 x^3+37x^2+12 x-6)^8\\[-0.1cm]
	& (9 x^4-11 x^3+15 x^2-8 x+4)^8 (9 x^4+29 x^3-15 x^2-28x+14)^8\\[-0.1cm] 
	& (9 x^4+80 x^3-12 x^2-136 x+68)^9 (10 x^4+9 x^3-103 x^2+188x-94)^9\\[-0.1cm] 
	&(12 x^4-85 x^3+115 x^2-60 x+30)^{10} (13 x^4-16 x^3+40 x^2-48x+24)^{11}\\[-0.1cm] 
	& (23 x^4+16 x^3-40 x^2+48 x-24)^{11} (26 x^4-83 x^3-19 x^2+204x-102)^{10}\\[-0.1cm] 
	& \underbrace{(60 x^4-79 x^3-215 x^2+588 x-294)^8 (x^6-6 x^5+11 x^4-8 x^3-x^2+6 x-2)^9}_{=F_{\text{nonlin}}(x)}
\end{align*}
with degree $1393$. Consequently, the defining recurrence for $h(n)$ has also degree $1393$ and one has to compute first $1393$ initial values to determine the remaining values efficiently by this recurrence. To derive these values, e.g., by a Taylor expansion is by far too slow. Thus we proceed differently by applying first partial fraction decomposition. Since also here the internal Mathematica version is too slow, we derived our own version by combining well known techniques in a clever way. First, we compute polynomials $s(x)$ and $t(x)$ such that the Bezout relation
\begin{equation}\label{Equ:BezoutRel}
	1=s(x)F_{\text{nonlin}}(x)+t(x)F_{\text{lin}}(x)
\end{equation}
holds. Here the extended Euclidean algorithm seems to be the first choice. Nevertheless, it turns out that solving the underlying linear system (related to the Sylvester matrix) is superior (by using efficient linear system solving). Using the found relation~\eqref{Equ:BezoutRel} and multiplying it with $\tfrac1{F_{\text{lin}}(x)\,F_{\text{nonlin}}(x)}$ yield the partial decomposition
$$\frac{1}{B(x)}=\frac{1}{F_{\text{lin}}(x)\,F_{\text{nonlin}}(x)}=\frac{s(x)}{F_{\text{lin}}(x)}+\frac{t(x)}{F_{\text{nonlin}}(x)}.$$
Next, we perform the classical method (by evaluation) to derive the partial fraction decomposition of $\frac{s(n)}{F_{\text{lin}}(n)}$. For the non-linear contribution we proceed as above by computing the corresponding Bezout relations via linear algebra. This finally leads to\footnote{Precisely these partial fraction decomposition techniques turned out to be useful in the summation package \texttt{Sigma}~\cite{Sigma} for simplifying sums~\cite{Schneider:23} that arise as solutions in linear difference equations~\cite{ABPS:21}.}
\begin{align*}
	\frac{1}{B(x)}=&\frac{a_1(x)}{(x-2)^{26}}+\frac{p_2(x)}{(x-1)^{19}}+\dots+\frac{p_{54}(x)}{(x^6-6 x^5+11 x^4-8 x^3-x^2+6 x-2)^9}.
\end{align*}
In this form we finally utilize the method of recurrences with constant coefficients (with order at most 6) to calculate the series expansions of each term in parallel and finally to get the coefficients $h(n)$ in~\eqref{Equ:1OverB}.
To this end, we can carry out the Cauchy products to get the expansions
\begin{align*}\frac{A(x)}{B(x)}\hat{I}(x)=&\underbrace{\underbrace{\Big(\sum_{n=0}^{\nu}p(n)x^n\Big)\Big(\sum_{n=0}^{\nu}h(n)x^n\Big)}_{\txt{Cauchy}}\Big(\sum_{n=0}^{\nu}I(n)x^n\Big)}_{\txt{Cauchy}}+O(x^{\nu+1})=\sum_{n=0}^{\nu}c(n)x^n+O(x^{\nu+1}).
\end{align*}
Also this task is challenging and further improvements can be applied as described, e.g. in~\cite{Blumlein:2023uuq}.

\section{The gluonic case}
\label{sec:gluonic}

We will now discuss the progress we have made so far in the gluonic case, which requires the calculation of a much larger number of expansion coefficients at $\hat{s}=0$.
We have been able to compute enough coefficients to obtain the recursion relations asscociated to the following constants
\begin{equation}
\left\{\zeta_2 \zeta_3, \, \zeta_5, \, \zeta_3, \, \ln(2) \zeta_2, \, \zeta_2^2, \, \ln^2(2) \zeta_2, \, {\rm Li}_4\left(\textstyle{\frac{1}{2}}\right), \, \ln^4(2) \right\},
\label{eq:constants3}
\end{equation}
of these, only the recursions associated to $\zeta_2 \zeta_3$ and $\zeta_5$ turned out to be solvable. In the other cases, a new feature appears compared with the quarkonic case:
There are new singularities in the differential equations, which forces us to also find power-log expansions at these singularities, as well as expansions around other points. 
Specifically, the singularities of the differential equations associated to
the constants $\zeta_2^2$, $\ln^2(2) \zeta_2$, ${\rm Li}_4\left(\textstyle{\frac{1}{2}}\right)$ and $\ln^4(2)$ are located at
\begin{equation}
\hat{s} = \left\{-4, -1, 3, 4\right\}.
\label{eq:singularities1}
\end{equation}
In the case of the constant $\ln(2) \zeta_2$, there are also singularities at $\hat{s}=-1/2$ and $\hat{s}=1$ in addition.
We will now take the opportunity to explain in more detail the different steps of the calculation.

\subsection{High energy expansion}

\begin{center}
\begin{figure}
\begin{center}
     \includegraphics[width=0.7\textwidth]{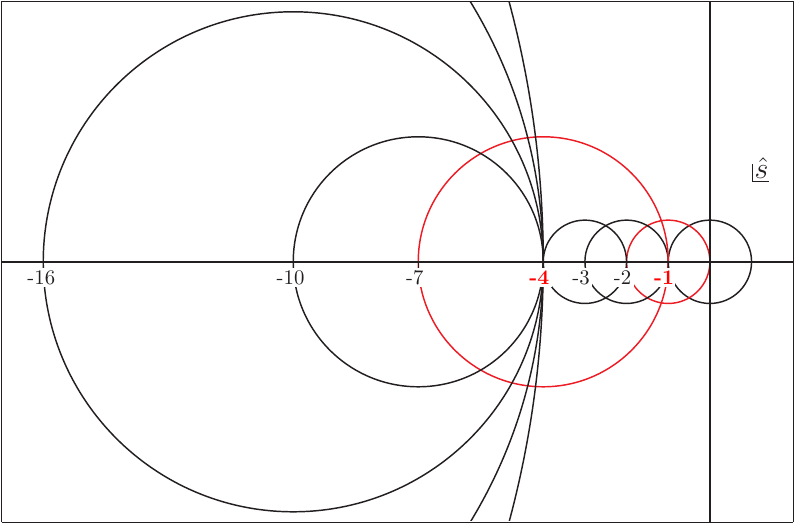}
\end{center}
\caption{\small\sf Radii of convergence of the successive points around which expansions are matched in the case of the last four constants in (\ref{eq:constants3}). 
The points $\hat{s}=-1$ and $\hat{s}=-4$, shown in red, are singularities of the differential equations.} 
\label{fig:convergence1a}
\end{figure}
\end{center}

As in the quarkonic case, we derived differential equations for the $F_{I;k,l}$'s, not only in the variable $\hat{s}$, but also in the variable $y = 1-x$,
from which we can then obtain the corresponding high energy expansions around $x=0$. 
In some cases, the indicial equation associated to the differential equation has only integer solutions, while in other cases, also half-integer solutions are present. 
In the former cases, we then have power-log expansions in the standard form
\begin{equation}
F_{I;k,l}(x) = \sum_{i=0}^2 \sum_{j=-2}^{\infty} c_{I;k,l}(i,j) x^j \ln^i(x)\, ,
\label{eq:x=0-expansion-1}
\end{equation}
while in the latter cases, we must use $\sqrt{x}$ as the expansion variable, so the expansions are given by
\begin{equation}
F_{I;k,l}(x) = \sum_{i=0}^2 \sum_{j=-2}^{\infty} c_{I;k,l}(i,j) (\sqrt{x})^j \ln^i(x).
\label{eq:x=0-expansion-2}
\end{equation}
After inserting either (\ref{eq:x=0-expansion-1}) or (\ref{eq:x=0-expansion-2}) in the differential equations in $x$, we obtain relations among the
coefficients $c_{I;k,l}(i,j)$, leaving only a few of them undetermined. 
As we explained in Section~\ref{sec:review}, these undetermined coefficients can then be obtained numerically using
matching conditions, see Ref. \cite{Blumlein:2023uuq} for more details. 
However, we cannot match the expansion at $x=0$ directly with the expansion at $\hat{s}=0$ ($x=1$), since, for example, in the case of the last four constants in (\ref{eq:constants3}),
the singular points $\hat{s}=-1$ and $\hat{s}=-4$ 
correspond in $x$ to $x=\frac{1}{2}\left(3-\sqrt{5}\right) \approx 0.38$ and $x=3-2 \sqrt{2} \approx 0.17$, respectively. In principle, we could try to get expansions around those two points in $x$,
but the corresponding differential equations blow up due to the square roots, and become impractical. We therefore find successive expansions at different points in $\hat{s}$, 
until we reach a value of $\hat{s}$ large enough that we can match the corresponding expansion with the one at $x=0$. 
We start by finding an expansion around the singular point $\hat{s}=-1$.
In the case of the four constants in (\ref{eq:constants3}) under consideration, this expansion will be of the form
\begin{equation}
F_{I;k,l}^{(-1)}(\hat{s}) = \sum_{j=-3}^{\infty} c_{I;k,l}^{(-1)}(0,j) z^j,
\label{eq:s=-1-expansion}
\end{equation}
where $z=\sqrt{1+\hat{s}}$. We see that in this case we do not get logarithms and the singularity is manifested by the negative powers in $z$. As we will be doing for all expansions, we find
a differential equation in $z$ from the one in $\hat{s}$,\footnote{One of the solutions to the indicial equation of this differential equation is $1/2$. Hence the square root in the change of
variables from $z$ to $\hat{s}$. This usually happens (although not always) around singular points.} then use this differential equation to find relations among the $c_{I;k,l}^{(-1)}(i,j)$'s 
and the compute the coefficients left undetermined by these relations by matching with the previous expansion (in this case, the one at $\hat{s}=0$) 
using a truncated version of the expansions (in this and other matchings, we used 5000 terms in the expansions, per power of the log, whenever logs were present).
The radius of convergence of this expansion (\ref{eq:s=-1-expansion}) is 1, which means we need to obtain an expansion around $\hat{s}=-2$, which will be of the form
\begin{equation}
F_{I;k,l}^{(-2)}(\hat{s}) = \sum_{j=0}^{\infty} c_{I;k,l}^{(-2)}(0,j) z^j,
\label{eq:s=-2-expansion}
\end{equation}
with $z=\hat{s}+2$. Again, the coefficients in (\ref{eq:s=-2-expansion}) are obtained by matching the expansions at $\hat{s}=-1$ and $\hat{s}=-2$, 
as we did with the expansions at $\hat{s}=0$ and $\hat{s}=-1$.
Notice the absence of logs or negative powers of $z$ in  (\ref{eq:s=-2-expansion}), which is expected since $\hat{s}=-2$ is not a singular point.
Since the closest singularity is at $\hat{s}=-1$ the radius of convergence of the expansion at $\hat{s}=-2$ is again 1. 
We then reapeat the process at $\hat{s}=-3$, where, again, the radius of convergence is 1 and the expansion is similar to (\ref{eq:s=-2-expansion}). 
We then match the expansion at $\hat{s}=-3$ with an expansion at $\hat{s}=-4$, which is of the form
\begin{equation}
F_{I;k,l}^{(-4)}(\hat{s}) = \sum_{i=0}^1 \sum_{j=-2}^{\infty} c_{I;k,l}^{(-4)}(i,j) z^j \ln^i(z),
\label{eq:s=-4-expansion}
\end{equation}
where $z=s+4$. After the point $\hat{s}=-4$, all expansions will be regular and similar to (\ref{eq:s=-2-expansion}). 
Since the nearest singularity is $\hat{s}=-1$, the radius of convergence of the expansion
(\ref{eq:s=-4-expansion}) is 3, so we can now match with an expansion at $\hat{s}=-7$. 
The radius of convergence of this expansion will also be 3, and we can then match it with an expansion at $\hat{s}=-10$,
which radius of convergence will be 6, and can therefore be matched with an expansion at $\hat{s}=-16$, and so on. 
We see that since the closest singularity for $\hat{s}<-4$ is the singularity at $\hat{s}=-4$, 
we can match expansions with radii of convergence that increase geometrically by a factor of two. 
We therefore do consecutive matchings of expansions at the following points:
\begin{equation}
\hat{s} = \left\{-196, \, -100, \, -52, \, -28, \, -16, \, -10, \, -7, \, -4, \, -3, \, -2, \, -1\right\},
\end{equation}
in reverse order. A few of these points, together with their corresponding convergence disks are shown in Fig. \ref{fig:convergence1a}. In the case of the constant $\ln(2) \zeta_2$, 
we also need to match at $\hat{s}=-1/2$ and $\hat{s}=-3/2$.
The point $s=-196$ corresponds in $x$ to $x = \left(99+70 \sqrt{2}\right)^{-1} \approx 0.005$, which is small enough to allow us to match this expansion with the expansion at $x=0$.
We did this for all cases available and found all coefficients left unconstrained by the differential equations with a precision of around a thousand digits. 
This was enough to obtain all coefficients using the PSLQ algorithm in terms of known constants, 
which in the cases under consideration were limited to the set $\left\{1,\pi,\zeta_2\right\}$, 
as expected since we are dealing at the moment with expansions associated to the weight 4 constants in (\ref{eq:constants3}), 
and therefore, the weight of any new constants on top of them cannot be higher than 2. 
Odd powers of $\pi$ only appear in the imaginary parts of the coefficients, which should cancel in the final result for the form factors.

\subsection{The $\hat{s}=4$ threshold}

Let us focus again on the last four constants in (\ref{eq:constants3}). Since the differential equations have singularities at $\hat{s}=3$ and $\hat{s}=4$, 
we again need to match consecutive expansions until we reach the $\hat{s}=4$ threshold. 
As we already mentioned, the radius of convergence of the expansion around $\hat{s}=0$ is 1, so we need to match with an expansion around $\hat{s}=1$. 
The radius of convergence of the latter expansion is now 2, since the closer singularity is at $\hat{s}=-1$, so in principle we could match this expansion directly with the
one at $\hat{s}=3$. However, we found better precision by matching first with an expansion around $\hat{s}=2$ before reaching $\hat{s}=3$. The expansions around $\hat{s}=3$ are of the form
\begin{equation}
F_{I;k,l}^{(3)}(\hat{s}) = \sum_{j=-3}^{\infty} c_{I;k,l}^{(3)}(0,j) z^j,
\label{eq:s=3-expansion}
\end{equation}
where $z=\sqrt{3-\hat{s}}$. We see that the singularity is manifested in terms of negative powers of $z$. However, it is interesting to notice that although the differential equations
allow us to search for such type of solutions, once we match with the expansion at $\hat{s}=2$, all coefficients of the negative powers in (\ref{eq:s=3-expansion}) turn out to be effectively zero, 
and therefore the singularity disappears.

\begin{center}
\begin{figure}
\begin{center}
\begin{minipage}[c]{0.45\linewidth}
     \includegraphics[width=1\textwidth]{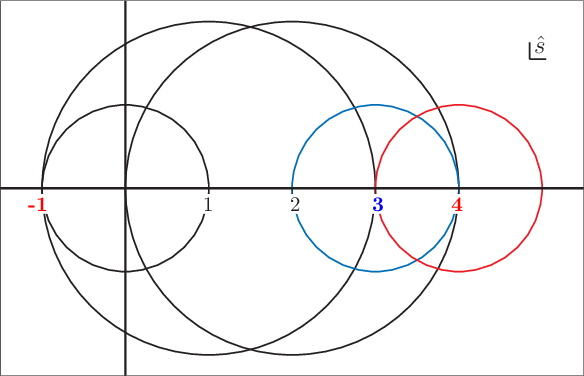}
\end{minipage}
\hspace*{0.01\linewidth}
\begin{minipage}[c]{0.45\linewidth}
     \includegraphics[width=1\textwidth]{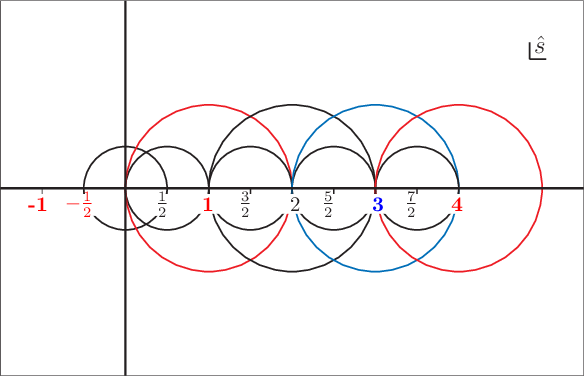}
\end{minipage}
\end{center}
\caption{\small\sf Radii of convergence of the successive points around which expansions are matched
in order to obtain the threshold expansion at $\hat{s}=4$ (shown in red). 
The point $\hat{s}=3$ (shown in blue) is singular according to the differential equations,
but the singularity disappears due to initial conditions. 
Left panel: radii of convergence of the points used in the case of the last four constants in (\ref{eq:constants3}).
Right panel: radii of convergence of the points used in the case of $\ln(2) \zeta_2$.} 
\label{fig:convergence2a}
\end{figure}
\end{center}

Finally, we match the expansions at $\hat{s}=3$ and $\hat{s}=4$. The latter has the following form
\begin{equation}
F_{I;k,l}^{(4)}(\hat{s}) = \sum_{i=0}^1 \sum_{j=-3}^{\infty} c_{I;k,l}^{(4)}(i,j) z^j \ln^i(z),
\label{eq:s=4-expansion}
\end{equation}
where $z=\sqrt{4-\hat{s}}$. All of the expansions from $\hat{s}=1$ to $\hat{s}=4$ were matched using 10000 expansion terms (per power of the log). 
These expansion points and their corresponding convergence disks are depicted in left panel of Fig. \ref{fig:convergence2a}. 
On the right panel, we show the corresponding convergence disks in the case of the constant $\ln(2) \zeta_2$, where we also have a singularity at $\hat{s}=1$.
In this case, in order to improve the precision of the expansion coefficients at $\hat{s}=4$, we matched successive expansions at every integer and haf-integer value of $\hat{s}$.
The corresponding radii of convergence are shown on the right panel of Fig. \ref{fig:convergence2a}.
We were able to determine the coefficients of the expansions at $\hat{s}=4$ with a precision of around 1400 decimal places. 
Again, we tried to use the PSLQ algorithm to try to express the coefficients in terms of known constants. Much like in the quarkonic case,
not all coefficients could be determined this way, but we could reduce the number of undetermined coefficients to just 5 in total in the case of the last four constants in (\ref{eq:constants3}).

\section{Conclusions}
\label{sec:conclusions}

We have made substantial progress applying the large moments method to the calculation of massive three-loop non-singlet QCD form factors. 
This method allows us to find recursions and differential equations associated to the constants appearing in the corresponding low energy expansions.
The recursion relations can then be used to obtain an even larger number of coefficients than the ones used to derive them, 
which allows us to get very precise numerical evaluations within the radius of convergence of the expansions. The differential equations can then be used to obtain expansions
at other points, which can be expressed in terms of just few of the coefficients. These coefficients can then be computed numerical by matching the expansions at
intermediate points. In this way, we were able to find expansions around all kinematic points of interest. 
In the gluonic case, this procedure turns out to be somewhat more complicated than in the quarkonic case,
since the differential equations have additional spurious singularities (they must cancel in the final physical result), which forces us to obtain successive expansions at many
intermediate points. So far we have been able to apply this method to five of the constants appearing the expansion at $\hat{s}=0$ ($x=1$), 
which required only 1000 coefficients in the case of the last four constants in (\ref{eq:constants3}), around 6000 coefficients in the case of $\ln(2) \zeta_2$,
and no more than 14500 coefficients in the case of $\zeta_3$. 
The remaining constants require a much larger number of coefficients, which are being computed at the moment. 
Once we obtain them, we will be able to proceed as in the other cases.

\vspace*{2mm}
\noindent
{\bf Acknowledgment.} 
This work has been supported in part  by the Austrian Science Fund (FWF) 10.55776/P33530.


\end{document}